\documentclass[aip,jcp]{revtex4}

\usepackage{graphicx}
\usepackage{tabularx}
\usepackage{color}
\usepackage{amsmath}
\usepackage{mathtools}
\usepackage{comment}
\newcommand{\be}{\begin{equation}}
\newcommand{\ee}{\end{equation}}
\newcommand{\bea}{\begin{eqnarray}}
\newcommand{\eea}{\end{eqnarray}}
\usepackage{dcolumn}
\usepackage{hyperref}
\usepackage{bm}
\usepackage{epsf}
\usepackage{subfigure}
\usepackage{epstopdf}%
\usepackage{amsfonts}
\usepackage{braket}
\usepackage{cancel}

\newcommand{\intinf}{\int_{-\infty}^{\infty}}

\begin{document}

\title{Qubit absorption refrigerator at strong coupling}  

\author{Anqi Mu$^1$, Bijay Kumar Agarwalla$^2$, Gernot Schaller$^3$, Dvira Segal$^1$}

\address{
$^1$Department of Chemistry and Centre for Quantum Information and Quantum Control, 
University of Toronto, 80 Saint George St., Toronto, Ontario, Canada M5S 3H6
}
\address{$^2$Department of Physics, Indian Institute of Science Education and Research,
Dr. Homi Bhabha Road, Pune 411 008, India}

\address{ $^3$Institut f\"ur Theoretische Physik, Technische Universit\"at Berlin, D-10623 Berlin, Germany}

\date{\today}

\begin{abstract}
We demonstrate that a quantum absorption refrigerator can be realized from the smallest quantum system, a qubit,
by coupling it in a non-additive (strong) manner to three heat baths.
This function is un-attainable for the qubit model under the weak system-bath coupling limit, when the dissipation is additive.
In an optimal design, the reservoirs are engineered and characterized by 
a single frequency component. We
obtain then closed expressions for the cooling window and refrigeration efficiency, as well as 
bounds for the maximal cooling efficiency and the efficiency at maximal power.
Our results agree with macroscopic designs and with three-level models for quantum absorption refrigerators,
which are based on the weak system-bath coupling assumption.
Beyond the optimal limit, we show with analytical calculations and numerical simulations 
that the cooling efficiency varies in a non-universal manner with model parameters.
Our work demonstrates that strongly-coupled quantum machines
can exhibit function that is un-attainable under the weak system-bath coupling assumption.
\end{abstract}

\maketitle

\section{Introduction}
\label{sec-intro}

An autonomous absorption refrigerator transfers thermal energy from a cold ($c$) bath to a hot ($h$) bath
without input power, by utilizing heat from an additional heat bath, a so-called work ($w$) reservoir.
Classical, large-scale absorption refrigerators were realized in the 19th century~\cite{AR19}, 
playing an important role in the development of the theory of irreversible thermodynamics. 
Proposals for quantum, nanoscale analogues of such machines aspire to establish the
theory of thermodynamics from quantum principles~\cite{review1,kos13,reviewARPC14}.

Quantum thermodynamical machines differ from their classical counterparts in two central aspects.
First, their performance relies on quantum phenomena
such as the 
discreteness 
of the energy spectrum of the working medium and quantum statistics.
Moreover, nontrivial quantum effects such as quantum coherence in the system,
\cite{scully0, scully1, scully2, HuberC,Anders} or in the bath~\cite{Lutz,Alicki15, Bijay-squeeze,manzano2016a}, 
quantum correlations~\cite{Huber,Alickientang}, non-locality, 
measurement~\cite{Kur,Alicki13}, and quantum driving and control~\cite{jacobs2009a,strasberg2013b}, 
may offer new principles for thermal machines. 
Beyond quantum resources, a second, fundamental aspect of nanoscale heat machines is that they
may operate beyond the weak system-bath coupling limit 
\cite{DavidS, Kosloff,Cao,Gernot,Eisert-strong,Jarzynski,esposito17,Nazir,Celardo}.
Classical-macroscopic thermodynamics is a weak-coupling theory; the impact of the
system-bath interface is small relative to the bulk behavior.
In contrast, small systems can strongly couple to their surroundings,
in the sense that the interaction energy between the system and the bath
becomes comparable to frequencies of the isolated system.

The goal of the present paper is to demonstrate that strongly-coupled system-bath quantum machines 
can exhibit function that is un-attainable under the weak coupling assumption.
We do so by analyzing additive and non-additive system-bath interaction models,
i.e., where the generator may or may not be additively decomposed into individual generators from the connected reservoirs.
In the additive case, the system (working medium) separately-independently exchanges energy with the hot, cold and work reservoirs.
In the non-additive model, 
the reservoirs inseparably interact with the system, thus acting
in a concerted-cooperative manner. 

Specifically, we show that a two-level system (TLS) 
cannot operate as an autonomous quantum absorption refrigerator (QAR) under the weak system-bath coupling approximation with
additive dissipators.
However, the same system does function as a QAR 
once it is allowed to couple to its surrounding reservoirs in a non-additive manner---representing strong coupling. 
Moreover, the qubit QAR can be optimized to perform 
at the maximal Carnot efficiency, and its performance is compatible with previous designs using three-level models, which
were constrained to operate under Lindblad dynamics with additive dissipators~\cite{jose13,joseSR,jose14}.
%
The smallest possible QAR described here relies on quantum principles and strong system-bath coupling effects. These
unique aspects are inherent to nanoscale devices.

This work is organized as follows:
We first introduce our model, showing that a QAR mode is impossible in the additive case (Sec.~\ref{subsec-add}), 
and afterwards present the non-additive model (Sec.~\ref{subsec-nonadd}), for which we present the basic definitions of energy currents
for two and three reservoirs.
Next, we present analytical results for the non-additive model, first on the cooling window and efficiency for specific spectral densities in
Sec.~\ref{subsec-cooling_efficiency} and then on the cooling efficiency at maximum power in Sec.~\ref{subsec-efficiency_max_power}.
For rectangular spectral densities we confirm these results by numerical simulations in Sec.~\ref{sec-simul}.
Finally, we show explicitly how such a non-additive dissipator may arise in the strong-coupling limit in Sec.~\ref{sec-physics}.

\begin{figure}[htbp]
\includegraphics[width=15cm]{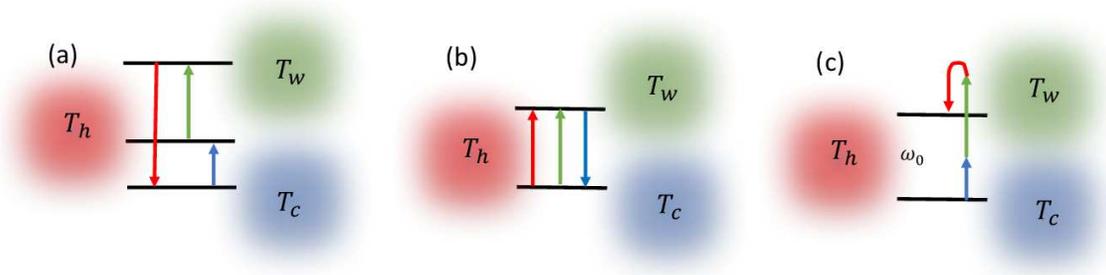}
\caption{(a) A three-level quantum system, with each transition thermalized separately with respect to an independent  
bath, hot (h), cold (c)  and work (w).
This model, which can operate as an absorption refrigerator~\cite{reviewARPC14, joseSR}, is not examined in the present work.
(b) A two-level system with an additive interaction model, Eq. ~(\ref{eq:H}). 
This system cannot act as a QAR under the weak-coupling approximation. 
(c) A two-level system with an inseparable interaction model to the baths, Eq.~(\ref{eq:Hs}). 
The concerted action of the baths is represented
by the arrows, acting together to e.g., excite the qubit. 
This model is examined in the present work. When optimized, it
performs as a QAR under conditions analogous to the three-level QAR of panel (a).
}
\label{FigQAR}
\end{figure}

\section{Model}
\label{sec-model}

A common design of an autonomous QAR consists of a three-level quantum system and 
three independent thermal reservoirs~\cite{reviewARPC14}.
Each transition between a pair of levels is {\it weakly} coupled to only one of the three
heat baths, $c$, $h$ and $w$, where  $T_w>T_h>T_c$, see Fig.~\ref{FigQAR}a.
In the steady state limit, the (ultra-hot) work bath provides energy to the system. This allows the extraction of energy
from the cold bath, to be dumped into the hot reservoir. The opposite
heating process, from the hot bath to the cold, takes place as well, but it can be minimized by manipulating the
frequencies of the system. 
The three-level QAR was discussed in details in several recent studies, see e.g. Refs.~\cite{reviewARPC14,joseSR,jose14}.
It is designed to perform optimally under the weak coupling approximation,
when each bath individually couples to a different transition. Quantum coherent and strong coupling effects are expected 
to reduce the cooling performance of a three-level QAR. 

In this paper we focus on a  QAR made of a {\it two-level system} coupled to three independent thermal baths.
When the baths couple to the qubit in an additive manner (Fig.~\ref{FigQAR}b)
we prove next that it is impossible to cool down the cold bath when the system-bath coupling is weak.
By allowing for  
cooperative system-bath interaction between the qubit and the reservoirs (Fig.~\ref{FigQAR}c), we are able to receive
a cooling condition, as well as derive bounds for the maximal efficiency of the QAR and its maximal power efficiency.



\subsection{Un-attainability of cooling for an additive dissipation model}
\label{subsec-add}

The additive model comprises a two-level system (spin, qubit) 
and three independent thermal reservoirs $\nu=c,h,w$,  $T_w>T_h>T_c$, $\beta_{\nu}=1/T_{\nu}$ with $k_B=1$.
The generic Hamiltonian is written as
%
\bea
\hat H = \frac{\omega_0}{2}\hat \sigma_z 
+\sum_{\nu}\hat H_{B,\nu} + 
\frac{\hat \sigma_x}{2} \otimes 
\left(\hat A_{c} + \hat A_h + \hat A_w\right).
\label{eq:H}
\eea
Here, $\hat \sigma$ are the Pauli spin matrices.
$\hat H_{B,\nu}$ is the Hamiltonian of the $\nu$-th reservoir. It includes, for example, a collection of
harmonic oscillators of frequencies $\omega_{j,\nu}$,
 $\hat H_{B,\nu}=\sum_{j}\omega_{j,\nu}\hat b_{j,\nu}^{\dagger}\hat b_{j,\nu}$
with $b^{\dagger}$ ($\hat b$) as bosonic creation (annihilation) operators.
The bath operator $\hat A_{\nu}$ is assumed to be hermitian. It couples the bath $\nu$ to the spin, where e.g.
$\hat A_{\nu}=\sum_{j}\lambda_{j,\nu}  \left(\hat b_{j,\nu}^{\dagger}+\hat b_{j,\nu}\right)$ 
with coupling strength $\lambda_{j,\nu}$.

Assuming a factorized-product initial state, $\langle \hat A_{\nu}\rangle=0$ with
the average performed over the initial-canonical state  of the bath,
weak system-bath coupling and Markovian dynamics,
we obtain a second order perturbative, Markovian quantum master equation~\cite{Breuer}.
This standard Born-Markov scheme results in
the stationary populations of the excited and ground state, respectively,
\bea
p_e=\frac{k_u}{k_d+k_u},\,\,\,\, p_g=\frac{k_d}{k_d+k_u},
\label{eq:pop}
\eea
with  $k_{d,u}=\sum_{\nu}k_{d,u}^{(\nu)}$.
The decay ($d$) and excitation ($u$) rate constants $k_{d,u}^{(\nu)}$,
induced by the $\nu$-th bath, depend on the details of the model.
The detailed balance relation dictates local thermal equilibrium, $k_u^{(\nu)}/k_d^{(\nu)}=e^{-\beta_{\nu}\omega_0}$.
The energy current, defined positive when flowing towards the qubit, can be similarly derived from the
Born-Markov approximation, and it is given by $J_c=-\omega_0 \left(  k_d^{(c)}  p_e - k_u^{(c)} p_g\right)$
\cite{segal-nitzan, segal-master, segal-nicolin, nazim}.
Substituting the steady state population~(\ref{eq:pop})  we obtain 
\bea
J_c
=-
\frac{\omega_0}{k_d+k_u}
\left[ k_d^{(c)}k_d^{(h)}  \left( \frac{k_u^{(h)}} {k_d^{(h)}} - \frac{k_u^{(c)}}{k_d^{(c)}}  \right) +
 k_d^{(c)}k_d^{(w)}   \left(\frac{k_u^{(w)}}{k_d^{(w)}} - \frac{k_u^{(c)}}{k_d^{(c)}}
\right) \right].
\label{eq:Jweak}
\eea
Using the detailed-balance relation and the fact 
that $(e^{-\beta_{h,w}\omega_0}-  e^{-\beta_c\omega_0})>0$, we conclude that $J_c<0$
irrespective of the details of the model.
Equation~(\ref{eq:Jweak}) reveals that under the additive model at weak coupling,
every two reservoirs exchange energy independently.
The prefactor in the denominator, $k_d+k_u$, which includes contributions from the three baths, 
only renormalizes the current. 
Since every two baths separately communicate, thermal energy always flows towards the colder bath,
and a chiller performance is un-attainable.

It should be pointed out that time-dependent, {\it driven} or {\it stochastic} models 
can realize refrigeration based on a qubit as a working medium even at weak coupling, 
see e.g.  Refs.~\cite{kos0,kos1,kos2,kos3,segal1,segal2,segal3,kur13}. 
These type of driven machines are beyond the scope of our work.

\subsection{Non-additive (strong) coupling model}
\label{subsec-nonadd}

It is evident that to realize a QAR  with a qubit as  the working substance, 
we must go beyond the model Hamiltonian~(\ref{eq:H}), or the weak-coupling approximation.
Our starting point is a revised  Hamiltonian with a {\it built-in} strong-coupling characteristic, 
a non-additive system-bath interaction operator,
\bea
\hat H = 
\frac{\omega_0}{2}\hat \sigma_z
+\sum_{\nu}\hat H_{B,\nu} +
\gamma \frac{\hat\sigma_x}{2} \otimes\left( \hat B_c \otimes \hat B_h \otimes \hat B_w\right).
\label{eq:Hs}
\eea
Here, $\hat B_{\nu}$ are bath operators, assumed to be hermitian, and $\gamma$ is an energy parameter characterizing the interaction energy.
The non-additivity of our model is assumed
to arise from  a more fundamental Hamiltonian with strong interactions between the quantum system
and individual reservoirs~\cite{segal-nitzan, segal-master,segal-nicolin}, see Sec.~\ref{sec-physics}.
%
%
Non-additive  models such as~(\ref{eq:Hs})
can be also accomplished by engineering many-body Hamiltonians based on e.g. resonant conditions and selection
rules. 

We emphasize that our model~(\ref{eq:Hs}) differs in a fundamental way from the QAR model analyzed theoretically
in e.g. Refs.~\cite{Levy12,reviewARPC14,plenio} and realized experimentally in a recent study~\cite{QAR-exp}. 
In Refs.~\cite{Levy12,reviewARPC14,plenio,QAR-exp},
the working medium includes three degrees of freedom such as three harmonic oscillators,
which interact via a three-body interaction term. 
Each oscillator is independently coupled to its own thermal bath, taken into account
by introducing additive Lindblad dissipators into the time evolution equation.
In contrast, in our model~(\ref{eq:Hs}) the quantum system is as simple as it can be, a qubit.
Nonlinearity is encoded into the model by assuming a non-additive interaction Hamiltonian with the three baths. 
This inseparability prevents us from arriving at standard
perturbative quantum master equations with additive dissipators 
(standard multi-terminal Lindblad or Redfield).


Back to Eq.~(\ref{eq:Hs}), we study the system's dynamics
assuming a fully factorized initial state by
using the Born-Markov approximation with the perturbative parameter $\gamma$.
While this is 
analogous to a weak coupling treatment, we emphasize again that the model is defined with an inherent
strong-coupling feature, the non-additivity of the interaction.

\subsubsection{Two-bath model}
\label{sec-two}
Equations of motion for the spin polarization, as well as the energy current, were derived in
Refs.~\cite{segal-nicolin, hava} for the model~(\ref{eq:Hs}) with two baths (hot and cold). 
The derivation relies on the assumption $\langle \hat B_{\nu}\rangle=0$, which could be satisfied exactly or
under conditions such as strong coupling or high temperature \cite{segal-nicolin}. Further, this assumption
can be relaxed by re-defining the model Hamiltonian to add and subtract the thermal average of the interaction Hamiltonian,
re-diagonalizing then the system's Hamiltonian and proceeding with the perturbative treatment  \cite{Ren1,Ren2}. 
The population dynamics~\cite{segal-nitzan, segal-master, segal-nicolin} satisfies
\bea
\dot p_e=-M(\omega_0) p_e(t)  + M(-\omega_0)p_g(t),
\label{eq:popd}
\eea
with rate constants
\bea
M(\omega_0)&=&\left(\frac{\gamma}{2}\right)^2\intinf e^{i\omega_0t} M_{h}(t) M_c(t) dt
\nonumber\\
& = & \frac{1}{2\pi} \left(\frac{\gamma}{2}\right)^2
\int_{-\infty}^\infty
M_h(\omega_0 - \omega)M_c(\omega) d\omega.
 \label{eq:conv} 
\eea
Here, $M_{\nu}(t)=\langle  \hat B_{\nu}(t) \hat B_{\nu}(0)\rangle$  is the two-time correlation function
with the average performed with respect to the canonical (initial) state of the $\nu$ thermal bath.
In Fourier space we introduce $M_{\nu}(\omega)=\intinf e^{i\omega t} M_{\nu}(t)dt$. 
In what follows,  we refer to this function as the ``Fourier bath-correlation function" (FBCF). 
This function
is real valued and positive.
In our work, the FBCF has a physical dimension of inverse energy ($\hbar=1$).
Formally similar to $P(E)$ theory~\cite{ingold1992a}, 
the detailed-balance condition is satisfied for the individual components,
$\frac{M_{\nu}(\omega) }{M_{\nu}(-\omega)} = e^{\beta_{\nu}\omega}$, but we do not have such a relation for 
the convoluted rate constant $M(\pm \omega_0)$.
%
%
Within the same treatment, the thermal energy current, flowing from the cold bath to the system, 
is given by a rather intuitive expression~\cite{segal-nicolin,hava},
\bea
J_c= -p_e \int_{-\infty} ^{\infty}d\omega \omega M_c(\omega)M_h(\omega_0-\omega)
+p_g \int_{-\infty} ^{\infty}d\omega \omega M_c(-\omega)M_h(\omega-\omega_0).
\label{eq:J1}
\eea
Here, we have absorbed a factor $C_2=\frac{1}{2\pi}\left(\frac{\gamma}{2}\right)^2$ in the definition of the current. 
%
The heat current  exhibits cooperative energy transfer processes:
The first term describes contributions to the current due to the decay of the qubit,
with $\omega$ energy exchanged with the cold bath and the rest $\omega_0-\omega$ absorbed or released by the hot bath.
A similar reasoning applies to the second term in Eq.~(\ref{eq:J1}),
which describes the excitation of the qubit.
%
For simplicity, in what follows we eliminate the spin gap and take $\omega_0=0$.
We then immediately conclude that $p_e=p_g=1/2$, thus  the heat current~(\ref{eq:J1}) simplifies to 
\bea
J_c&=& -\frac{1}{2}\int_{-\infty} ^{\infty}d\omega \omega M_c(\omega)M_h(-\omega)
+ \frac{1}{2}\int_{-\infty} ^{\infty}d\omega \omega M_c(-\omega)M_h(\omega)
\nonumber\\
&=&\frac{1}{2}
\int_{-\infty}^{\infty}d\omega \omega M_c(\omega)M_h(\omega)\left[e^{-\beta_c\omega} - e^{-\beta_h\omega}\right]
\nonumber\\
&=&  \int_{-\infty}^{\infty}d\omega \omega M_c(-\omega)M_h(\omega).
\eea
%
Eq.~(\ref{eq:J1}) can be derived from a full counting statistics perspective, and
the resulting cumulant generating function satisfies the steady state heat exchange fluctuation theorem 
\cite{segal-nicolin,Ren1,Ren2,hava}.
Therefore, our framework satisfies the second law of thermodynamics, which is a direct consequence of the exchange fluctuation
relation.
From this, it is easy to prove that if $\beta_c>\beta_h$, $J_c<0$, meaning that the heat current flows towards the cold bath.

\subsubsection{Three-bath case}
Back to the QAR model Hamiltonian~(\ref{eq:Hs}), a full-counting statistics analysis allows us to
describe energy exchange with three thermal reservoirs~\cite{hava}, in a complete analogy to the two-bath case 
described in Sec.~\ref{sec-two}. This formalism directly provides both population dynamics and the dynamics of
the so-called cumulant generating function, handing over all current cumulants.
The population dynamics follows Eq.~(\ref{eq:popd}), with the FBCF 
now however given by
\bea
M(\omega_0)&=&\left(\frac{\gamma}{2}\right)^2\intinf \intinf e^{i\omega_0t} M_{c}(t) M_h(t)M_w(t) dt
\nonumber\\
& = & \frac{1}{(2\pi)^2} \left(\frac{\gamma}{2}\right)^2
\int_{-\infty}^\infty \intinf
M_c(\omega_1)
M_h(\omega_2)
M_w(\omega_0 - \omega_1-\omega_2) d\omega_1d\omega_2.
 \label{eq:conv3} 
\eea
The energy current, from the cold bath towards the qubit, is given by
\bea
J_c&=& - p_e\int_{-\infty} ^{\infty}\int_{-\infty} ^{\infty}d\omega_1 d\omega_2 
\omega_1 M_c(\omega_1)M_h(\omega_2)M_w(\omega_0-\omega_1-\omega_2)
\nonumber\\
&&+p_g \int_{-\infty} ^{\infty}\int_{-\infty}^{\infty} 
d\omega_1 d\omega_2 \omega_1 M_c(-\omega_1)M_h(-\omega_2) M_w(\omega_1+\omega_2-\omega_0).
\label{eq:Jc3}
\eea
Again, we have absorbed a prefactor $C_3=\frac{1}{(2\pi)^2}\left(\frac{\gamma}{2}\right)^2$ into the definition of the current. 
This intuitive expression, which can be also suggested phenomenologically~\cite{segal08},
describes coordinated three-bath energy exchange processes,
with an overall conservation of energy. An amount of energy $\omega_1$ is delivered to the cold bath 
or absorbed from it, while the other reservoirs assist by providing or absorbing the rest of the energy, so 
as to complete a decay (first integral) or an excitation (second line) process.
When $\omega_0=0$, we receive
%
\bea
J_c&=&
-\frac{1}{2}\int_{-\infty} ^{\infty}d\omega_1 \omega_1 M_c(\omega_1) \int_{-\infty}^{\infty}
d\omega_2 M_h(\omega_2) M_w(\omega_1+\omega_2)\left[ e^{-\beta_w(\omega_1+\omega_2)}- e^{-\beta_c\omega_1}e^{-\beta_h\omega_2}
\right]
\nonumber\\
&=&
-\int_{-\infty} ^{\infty}\int_{-\infty} ^{\infty}d\omega_1 d\omega_2   
\omega_1 M_c(\omega_1)M_h(\omega_2)M_w(-\omega_1-\omega_2).
\label{eq:JQARu}
\eea
While we write here an expression for $J_c$ only, the full-counting statistics approach~\cite{hava} readily 
hands over analogous expressions for $J_w$ and $J_h$. Particularly, $J_h$ is received from 
Eq.~(\ref{eq:Jc3}) by interchanging the `c' and `h' indices, 
and $J_w$ then follows from energy conservation.

\section{Analytical Results}
\label{sec-anal}

\subsection{Cooling window and efficiency}\label{subsec-cooling_efficiency}
Can we realize a QAR based on a qubit, using the model~(\ref{eq:Hs})?
Our objective is to derive a cooling condition from Eq.~(\ref{eq:JQARu}), i.e., find out whether we can engineer the
system and the baths to achieve refrigeration, $J_c\geq0$. 

The FBCF $M_{\nu}(\omega)$ is related to the spectral density of the $\nu$ thermal bath, see Sec.~\ref{sec-physics}.
Let us assume that the baths are engineered such that these functions
are characterized by the frequency $\theta_{\nu}>0$, satisfying
a resonant assumption 
\bea
\theta_c+\theta_w =\theta_h.
\label{eq:resonant}
\eea
We now analyze the performance of our system as a QAR in an idealized limit, then under more practical settings.
The resonant assumption 
will be assumed throughout, 
though it is not a necessary condition for refrigeration in non-ideal designs as we demonstrate through simulations and discuss
in our conclusions.

\subsubsection{Ideal design}
%
In an optimal design, the FBCF $M_{\nu}(\omega)$  is restricted to be nonzero within a narrow spectral window.
In the most extreme case,  we filter all frequency components besides the central mode, 
\bea
M_{\nu}(\omega)=
\begin{cases}
\epsilon_{\nu}\delta_{\nu}(\omega-\theta_{\nu}) &  \text{for }  \omega\geq0\\
\epsilon_{\nu}\delta_{\nu}(\omega+\theta_{\nu}) e^{\beta_\nu \omega} 
&  \text{for } \omega\leq0
\end{cases}
\label{eq:Mdelta}
\eea
Here, $\epsilon_{\nu}>0$ is a dimensionless parameter.
We evaluate the integrals in Eq.~(\ref{eq:JQARu}), and find that there are only two contributions to the cooling current,
when $\omega_1\leq0$ and $\omega_2\geq0$, and the other way around.
We then receive the cooling condition, $J_c\geq 0$, 
\bea
\theta_c \left( e^{-\beta_c\theta_c} e^{-\beta_w\theta_w} - e^{-\beta_h\theta_h}\right)  \geq 0.
\label{eq:etadelta}
\eea
The first term describes the removal of heat from the cold bath---assisted by the work reservoir---and its release to the hot bath.
The second term accounts for the reverse process, with energy absorbed from the hot bath and released into both the cold and 
work reservoirs. Equation~(\ref{eq:etadelta}) can be also organized as follows,
\bea
\left(\frac{T_w-T_h}{T_w-T_c}\right)\frac{T_c}{T_h}
\geq \frac{\theta_c}{\theta_h},
\label{eq:cool-op}
\eea
which precisely corresponds to the cooling condition as obtained for a three-level QAR---analyzed with a Markovian 
master equation with additive dissipators~\cite{reviewARPC14,joseSR}.

\subsubsection{Non-ideal design}
Let us now consider a more physical---and less optimal model. We engineer the FBCFs as follows ($\omega>0$),
\bea
M_c(\omega)&=&\epsilon_c\delta(\omega-\theta_c),
\nonumber\\
M_h(\omega)&=&\epsilon_h \left[ H(\omega-\theta_h+\delta) - H(\omega-\theta_h-\delta) \right],
\nonumber\\
M_w(\omega)&=&\epsilon_w \left[ H(\omega-\theta_w+\delta) - H(\omega-\theta_w-\delta) \right].
\label{eq:steps}
\eea
For negative frequencies, the detailed balance relation multiplies each function by a thermal factor.
Here, $H(x)$ is the Heaviside step function, $\delta(x)$ is the Dirac Delta function
and $\epsilon_{\nu}$ is a dimensionless parameter. As we show below, these parameters, 
which characterize the reservoirs' spectral functions,  neither affect
the cooling window nor the efficiency. 
We assume the resonant assumption~(\ref{eq:resonant})
to hold, and that the width parameter $\delta$ is small, $\delta \ll \theta_{\nu}$. 

We calculate the cooling current based on Eq.~(\ref{eq:JQARu}) 
by breaking it into four contributions. Let us first inspect the 
$\omega_1\geq0$ and $\omega_2\geq 0$ term,
\bea
J_c^{(+,+)}
&=&
-\int_{0}^{\infty} d\omega_1 \omega_1 M_c(\omega_1) 
\int_{0}^{\infty} d\omega_2 M_h(\omega_2) M_w(\omega_1+\omega_2) e^{-\beta_w(\omega_c+\omega_h)}
\nonumber\\
&=&
-\theta_c\epsilon_c\epsilon_h
\int_{\theta_h-\delta}^{\theta_h+\delta} d\omega_2M_w(\theta_c+\omega_2) e^{-\beta_w(\theta_c+\omega_h)}.
\label{eq:JQARpp}
\eea
Since $\theta_c+\theta_h-\delta > \theta_w+\delta$, the integral collapses to null.
A similar argument brings a zero contribution from the negative branch, $J_c^{(-,-)}=0$, which includes
$\omega_1\leq0$ and $\omega_2\leq 0$.
We proceed and evaluate the contribution to the cooling current from $\omega_1\leq0$ but $\omega_2\geq0$. 
Recall that $\theta_h-\delta>\theta_c$, 
\bea
J_c^{(-,+)}
&=&
\int_{0}^{\infty} d\omega_1 \omega_1 M_c(\omega_1)e^{-\beta_c\omega_1} 
\int_{0}^{\infty} d\omega_2 M_h(\omega_2) M_w(-\omega_1+\omega_2) e^{-\beta_w(-\omega_1+\omega_2)}
\nonumber\\
&=&\epsilon_c
\theta_c e^{-\beta_c\theta_c} 
\int_{0}^{\infty} d\omega_2 M_h(\omega_2)M_w(\omega_2-\theta_c)e^{-\beta_w(\omega_2-\theta_c)}
\nonumber\\
&=& 
\epsilon_c\epsilon_h\theta_c  e^{-\beta_c\theta_c} 
\int_{\theta_h-\delta}^{\theta_h+\delta} 
d\omega_2 M_w(\omega_2-\theta_c)  e^{-\beta_w(\omega_2-\theta_c)} 
\nonumber\\
&=&
\frac{2 \epsilon_c \epsilon_h \epsilon_w}{\beta_w}\theta_c e^{-\beta_c\theta_c} e^{-\beta_w\theta_w} \sinh(\beta_w\delta).
\label{eq:JQARmp}
\eea
We had utilized the resonant assumption~(\ref{eq:resonant}) in the last line.
A similar analysis gives
\bea
J_c^{(+,-)}= 
-\frac{2\epsilon_c \epsilon_h \epsilon_w}{\beta_h}\theta_c e^{-\beta_h\theta_h} \sinh(\beta_h\delta).
\label{eq:JQARpm}
\eea
Putting together Eqs.~(\ref{eq:JQARmp}) and~(\ref{eq:JQARpm}), we organize the cooling condition, $J_c\geq 0$,  as
\bea
\frac{1}{\beta_w}\theta_c e^{-\beta_c\theta_c} e^{-\beta_w\theta_w} \sinh(\beta_w\delta)
-\frac{1}{\beta_h}\theta_c e^{-\beta_h\theta_h} \sinh(\beta_h\delta) \geq 0.
\eea
We Taylor-expand this result to the first nontrivial order in $\delta$, which is the third order, and receive
\bea
\left(2\delta +\frac{1}{3}\beta_w^2\delta^3\right) 
e^{-\beta_c\theta_c} e^{-\beta_w\theta_w} -
\left(2\delta +\frac{1}{3}\beta_h^2\delta^3\right)
e^{-\beta_h\theta_h} \geq 0.
\eea
Using $\theta_w=\theta_h-\theta_c$, we find that
\bea
(\beta_h-\beta_w)\theta_h - (\beta_c-\beta_w)\theta_c \geq \ln \frac{2\delta +\frac{1}{3}\beta_h^2\delta^3 }  
{  2\delta +\frac{1}{3}\beta_w^2\delta^3}.
\eea
Re-arranging this expression, we get a cooling condition
\bea
\left(\frac{T_w-T_h}{T_w-T_c}\right)\frac{T_c}{T_h}\geq \frac{\theta_c}{\theta_h}
-\frac{T_cT_w}{\theta_h(T_w-T_c)}
\ln \left[\frac{2+\frac{{\beta_w}^2\delta^2}{3}}{2+\frac{{\beta_h}^2\delta^2}{3}}\right].
\label{eq:cool1}
\eea
%
In the  limit $\delta \rightarrow 0$,  we retrieve Eq.~(\ref{eq:etadelta}), 
which corresponds to the three-level QAR analyzed with a Markovian master equation  with additive dissipators
~\cite{reviewARPC14,joseSR}.
It is important to recognize that the $\delta$ dependence in Eq.~(\ref{eq:cool1}) is non-universal, and it depends on
our choice~(\ref{eq:steps}).
For example, using a different setting,  with $M_{c}(\omega)$ and $M_h(\omega)$ as step functions of
width $2\delta$, but $M_w(\omega)$ a Dirac delta function,
we derive an alternative cooling condition,
\bea
\left(\frac{T_w-T_h}{T_w-T_c}\right)\frac{T_c}{T_h}  \geq
\frac{\theta_c}{\theta_h}
-\frac{T_cT_w}{\theta_h(T_w-T_c)}
\label{eq:cool0}
\ln\left[\frac{2\theta_c+\frac{1}{3}\theta_c{\beta_{c}}^2{\delta}^2-\frac{2}{3}{\beta_{c}}{\delta}^2}{2\theta_c+\frac{1}{3}\theta_c{\beta_{h}}^2{\delta}^2-\frac{2}{3}{\beta_{h}}{\delta}^2}\right].
\label{eq:cool2}
\eea
The role of the width parameter $\delta$ is non-trivial, and it can reduce or increase the cooling window. 

Arriving at equation~(\ref{eq:cool-op}), and receiving its generalizations to finite width,
Eqs.~(\ref{eq:cool1}) and~(\ref{eq:cool2}), are central results of our work.
The cooling window depends on $\delta$ in a non-universal manner.
In the optimal limit, $\delta \rightarrow 0$, we recover
the three-level weak-coupling condition, which is bounded by the Carnot limit of a macroscopic 
absorption refrigerator, 
as we show next.

The efficiency of a refrigerator is defined by the coefficient of performance (COP), the ratio
between the heat current removed from the cold bath  $J_c$ and the input heat from the work bath $J_w$, 
$\eta \equiv \frac{J_c}{J_w}$. For convenience, in what follows, we sometimes refer to the cooling COP  as the 
``efficiency" of the refrigerator,
in the sense that this measure characterizes the competence of the  machine, 
but remind the reader that it can e.g. assume values larger than one.

It is convenient to write down an equation for $J_h$,  analogous to Eq.~(\ref{eq:JQARu}), and evaluate it 
with the model~(\ref{eq:steps}).  The heat current from the work bath is given by $J_w=-J_c-J_h$.
The currents from the work and cold baths are
\bea
J_w&=& \frac{\epsilon_c\epsilon_h\epsilon_w}
{\beta_h}e^{-\beta_h\theta_h}
\left[\left(\theta_w+\delta+\frac{1}{\beta_h}\right)e^{-\beta_h\delta}
-\left(\theta_w-\delta+\frac{1}{\beta_h}\right)e^{\beta_h\delta}\right]
\nonumber\\
&&-
\frac{ \epsilon_c\epsilon_h\epsilon_w}{\beta_w}e^{-\beta_c\theta_c} e^{-\beta_w\theta_w}
\left[\left(\theta_w+\delta+\frac{1}{\beta_w}\right)e^{-\beta_w\delta}
-\left(\theta_w-\delta+\frac{1}{\beta_w}\right)e^{\beta_w\delta}\right],
\nonumber\\
J_c&=&\theta_c \epsilon_c\epsilon_h\epsilon_w \left\{
e^{-\beta_c\theta_c}
e^{-\beta_w\theta_w}
\left[e^{-\beta_w\delta}-e^{\beta_w\delta}\right]
\frac{1}{-\beta_w}+\frac{1}{\beta_h}e^{-\beta_h\theta_h}
\left[e^{-\beta_h\delta}-e^{\beta_h\delta}\right] \right\}.
\label{eq:Jhc}
\eea
We expand the currents to third order in $\delta$
and receive the cooling COP 
\bea
\eta=
\frac{\theta_ce^{-\beta_c\theta_c}
e^{-\beta_w\theta_w}[2\delta+\frac{1}{3}{\beta_w}^2{\delta}^3]-\theta_ce^{-\beta_h\theta_h}
\left[2\delta+\frac{1}{3}{\beta_h}^2{\delta}^3\right]}{e^{-\beta_c\theta_c}e^{-\beta_w\theta_w}
\left[2\theta_w\delta-\frac{2}{3}\beta_w{\delta}^3+
\frac{1}{3}\theta_w{\beta_w}^2{\delta}^3\right]-e^{-\beta_h\theta_h}
\left[2\theta_w\delta-\frac{2}{3}\beta_h{\delta}^3+
\frac{1}{3}\theta_w{\beta_h}^2{\delta}^3\right]}.
\label{eq:coolCOP}
\eea
The width parameter $\delta$ could affect the efficiency in a non-monotonic way.  
However, to the lowest (linear) order in $\delta$ we gather 
\bea
\eta= \frac{\theta_c}{\theta_w}.
\label{eq:etadelta0}
\eea
The cooling condition~(\ref{eq:cool-op})  can be used to derive a bound on efficiency. In the ideal limit,
the cooling window is defined from
\bea
\left(\frac{T_w-T_h}{T_w-T_c}\right)\frac{T_c}{T_h}
\geq \frac{\theta_c}{\theta_h} =  \frac{\theta_c}{\theta_c+\theta_w}.
\eea
We inverse this expression and reach
\bea
\left(\frac{T_w-T_c}{T_w-T_h}\right)\frac{T_h}{T_c} \leq 1+ \frac{\theta_{w}}{\theta_c},
\eea
which can be also expressed as
\bea
\left(\frac{T_h-T_c}{T_w-T_h}\right)\frac{T_w}{T_c} \leq \frac{\theta_{w}}{\theta_c},
\eea
finally receiving
$\frac{\theta_c}{\theta_w} \leq \left(\frac{T_w-T_h}{T_h-T_c} \right)\frac{T_c}{T_w}$.
Comparing this expression to Eq.~(\ref{eq:etadelta0}), we gain a bound on the cooling COP (in the $\delta \rightarrow 0$ limit),
\bea
\eta= \frac{\theta_c}{\theta_w} \leq \left(\frac{T_w-T_h}{T_h-T_c} \right)\frac{T_c}{T_w}\equiv\eta_c,
\label{eq:eta}
\eea
which is nothing but the Carnot bound.
Specifically, when $T_w\gg T_c,T_h$, we  find that $\frac{\theta_c}{\theta_w}< T_c/(T_h-T_c)$, which is
the Carnot bound for cooling machines.
We emphasize that Eq.~(\ref{eq:eta}) was received in previous studies, see e.g. Ref.~\cite{joseSR}, yet restricted
to quantum systems that evolve with additive dissipators.
The COP of non-ideal models with a finite width $M_{\nu}(\omega)$, can be also calculated and simulated, as we do in Sec. 
\ref{sec-simul}. It is important to note that the COP of our model is always limited by the Carnot bound~(\ref{eq:eta})
as was recently proven in Ref.~\cite{Bijay-squeeze} based on the heat exchange fluctuation theorem.

The analysis presented in this subsection can be extended to describe a non-degenerate (biased) qubit $\omega_0\neq0$ 
as we show in Appendix A.
We find that for small $\omega_0$ we can adjust the functions $M_{\nu}(\omega)$  so as to
recover the ideal cooling window~(\ref{eq:etadelta}) and the corresponding efficiency bounds.

Another natural question concerns the potential to realize a QAR with
continuous functions $M_{\nu}(\omega)$, rather than the hard-cutoff model~(\ref{eq:steps}).
In Appendix B we calculate the
cooling current~(\ref{eq:JQARu}) using Gaussian functions for $M_{\nu}(\omega)$, and prove
that the system cannot act as a QAR.
It would be interesting to find whether there is a general proof for the un-attainability of cooling
once the functions $M_{\nu}(\omega)$ fully overlap.

\subsection{Efficiency at maximum power}\label{subsec-efficiency_max_power}

The discussion above concerns a bound for the maximal COP-obtained for vanishing cooling power.
For practical purposes, however, one needs to operate machines
at non-vanishing output power. A more practical figure-of-merit is the maximal power
efficiency (MPE) $\eta^*$, meaning, that we calculate the cooling COP when maximizing the cooling current.
An intriguing question, which  recently received much attention, is
whether $\eta^*$ can approach $\eta_c$--- albeit at finite cooling power. 
For heat engines, bounds for the MPE were extensively investigated~\cite{curzon1975,broeck,udo,esp1,esp2,yun,casatiRev}.
The MPE of refrigerators was examined in e.g., Ref.~\cite{joseSR}, demonstrating that it is 
constrained by the spectral properties of the thermal reservoirs. 
Nevertheless, Ref.~\cite{joseSR} 
was concerned with additive (weakly-coupled) system-bath models.

To derive an expression for the efficiency at           
maximal cooling power, we go back to Eq.~(\ref{eq:Jhc}), and analyze it
to the lowest order in $\delta$,
%
\bea
J_c&=&
2\delta \times \epsilon_c\epsilon_h\epsilon_w 
\theta_c \left( e^{-\beta_c\theta_c} e^{-\beta_w\theta_w} -  e^{-\beta_h\theta_h} \right)
\nonumber\\
&=& 2\delta \times \epsilon_c\epsilon_h\epsilon_w
\theta_c \left( e^{(\beta_w-\beta_c)\theta_c} e^{-\beta_w\theta_h} -  e^{-\beta_h\theta_h} \right).
\eea
This function is linear in $\theta_c$ for small values, but it decays exponentially with $\theta_c$ beyond that.
We thus search for $\theta_c$ which maximizes $J_c(\theta_c)$ 
by solving  $\frac{\partial J_c}{\partial \theta_c}=0$. We find that
\bea
\ln \left[1+\theta_c(\beta_w-\beta_c)\right] 
+ (\beta_w-\beta_c) \theta_c - \beta_w\theta_h = -\beta_h\theta_h.
\label{eq:eq}
\eea
Since $\ln(1+x)\leq x$, we turn the equality~(\ref{eq:eq}) into an inequality,
\bea
(\beta_w-\beta_h)\theta_h\leq 2(\beta_w-\beta_c)\theta_c,
\eea
or
\bea
\theta_c (2\beta_c-\beta_w-\beta_h) \leq \theta_w (\beta_h-\beta_w).
\eea
The MPE is thus bounded by 
\bea
\eta^*=\frac{\theta_c}{\theta_w} \leq \frac{\beta_h-\beta_w}{2\beta_c-\beta_w-\beta_h} = \frac{\eta_c}{\eta_c+2}.
\eea
Since $\beta_w\leq \beta_h$, we receive the bound
\bea
\eta^*\leq \frac{1}{2} \frac{\beta_h-\beta_w}{\beta_c-\beta_h} = \frac{1}{2}\eta_c
\eea
Interestingly, the efficiency at maximum cooling power is upper-bounded by half of the Carnot bound for cooling.
%
The second order term in the expansion is $-\frac{1}{4}\eta_c^2$, which is different than the value obtained 
for low dissipation engines
\cite{esp2}. 

We repeat this exercise using functions $M_{\nu}(\omega)$ of a finite width $\delta$. 
For example, we start from equation~(\ref{eq:Jhc}), expand it to third order in $\delta$, and solve 
$\frac{\partial J_c}{\partial \theta_c}=0$ to obtain a relation for $\theta_c$ which maximizes the cooling current.
Using again the fact that $e^x\geq 1+x$, we receive
\bea
\frac{\theta_c}{\theta_w} 
\leq \frac{1}{2} \eta_c + \frac{1} {\theta_w} \frac{1}{2\beta_c-\beta_w-\beta_h} 
\ln\left[\frac{2 +\frac{1}{3}\beta_w^2\delta^2 }{2+\frac{1}{3}\beta_h^2\delta^2} \right].
\label{eq:MPE}
\eea
%
The left hand side in this inequality corresponds to the COP, even at large $\delta$, as long as 
we work in the ``linear response limit" of $\beta_h\rightarrow \beta_w\rightarrow 0$,
see Eq. (\ref{eq:coolCOP}). The right hand side then collapses to $\eta_c/2$.
We therefore conclude  that in the linear response limit
the maximum power efficiency is upper bounded  by $\eta_c/2$, independent of the details of the model.

\section{Simulations}
\label{sec-simul}

In this section, we examine the performance of the qubit-QAR beyond the ideal (or close to ideal) limit as explored
in Sec.~\ref{sec-anal}. We perform simulation directly by using Eqs.~(\ref{eq:Jc3}) and the analogous expression for $J_w$.
For the FBCF we use the following model ($\omega\geq0$),
\bea
M_{\nu}(\omega)=
\omega^p \left[n_{\nu}(\omega)+1\right] \left[ H(\omega-\theta_{\nu}+\delta_{\nu}) - H(\omega-\theta_{\nu}-\delta_{\nu}) \right].
\label{eq:Msimul}
\eea
Here, $\delta_{\nu}$ is not necessarily small,
$n_{\nu}(\omega)=\left[\exp(\beta_{\nu}\omega)-1\right]^{-1}$ is the Bose-Einstein distribution function,
$H(x)$ is the Heaviside step function, 
and $p$ is a dimensionless parameter.
The same form holds for negative frequencies, but with a detailed balance factor such that
$\frac{M_{\nu}(\omega) }{M_{\nu}(-\omega)} = e^{\beta_{\nu}\omega}$.
The model is motivated from physical grounds in Sec.~\ref{sec-physics}.
In simulations we use $p=1$, thus one should add a prefactor
with units of an inverse energy to~(\ref{eq:Msimul}). This prefactor is taken as unity here.
We found that results did not qualitatively change when using other powers  $p$.
We assume the resonance assumption~(\ref{eq:resonant}) to hold, though this is not a necessary condition for realizing a qubit-QAR 
once we operate the system beyond the ideal limit (large $\delta_{\nu}$). 

\begin{figure}[htbp]
\hspace{4mm}
\includegraphics[width=8cm]{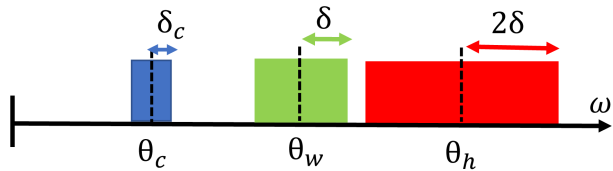}
\caption{Scheme of the spectral windows assumed in Fig.~\ref{Figcool} for
the simulation of the cooling current and the cooling COP.
We vary $\beta_h$ and $\delta$ fixing $\beta_c=1$, $\beta_w=0.1$,
$\theta_c=2$, $\theta_h=6$, $\theta_w=4$, $\omega_0=0$, $\epsilon_{\nu}=1$ 
We use the form~(\ref{eq:Msimul}) with $p=1$,
 $\delta_c=0.2$, $\delta_w=\delta$ and $\delta_h=2\delta$.
The rectangular boxes represent the windows over which the functions
$M_{\nu}(\omega)$ are defined---in the domain of positive frequencies. 
Corresponding functions appear in the negative range, decorated by a detailed balance factor 
$e^{\beta_\nu \omega}$.
}
\label{Figbox}
\end{figure}

As a concrete example, we assume that the cold bath has a finite-fixed  width of 
$\delta_c=0.2$, 
while the other two windows are tuned with $\delta_w=\delta$ and $\delta_h=2\delta$.
For a schematic representation, see Fig.~\ref{Figbox}.

\begin{figure}[htbp]
\hspace{4mm}
\includegraphics[width=7cm]{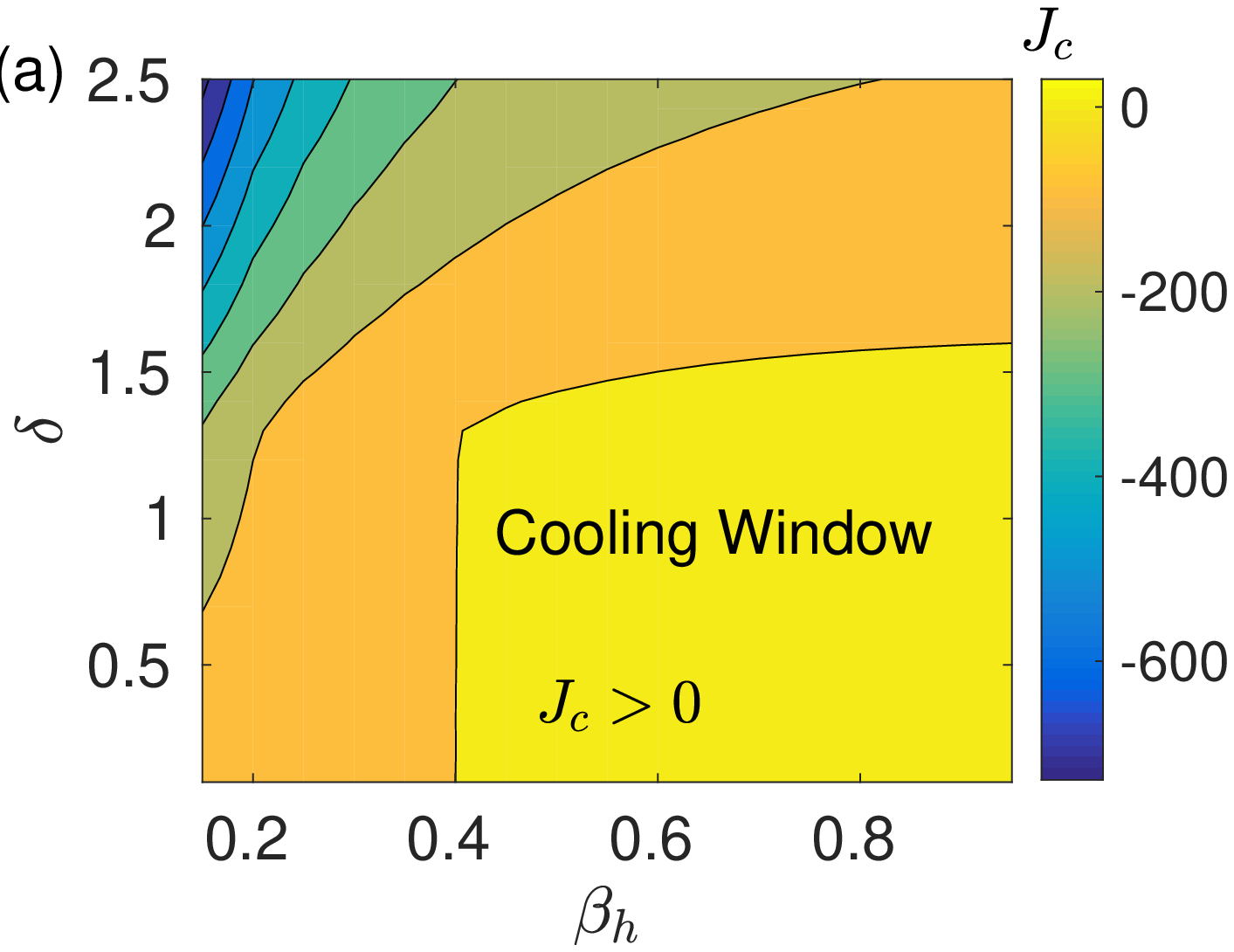}
\includegraphics[width=7cm]{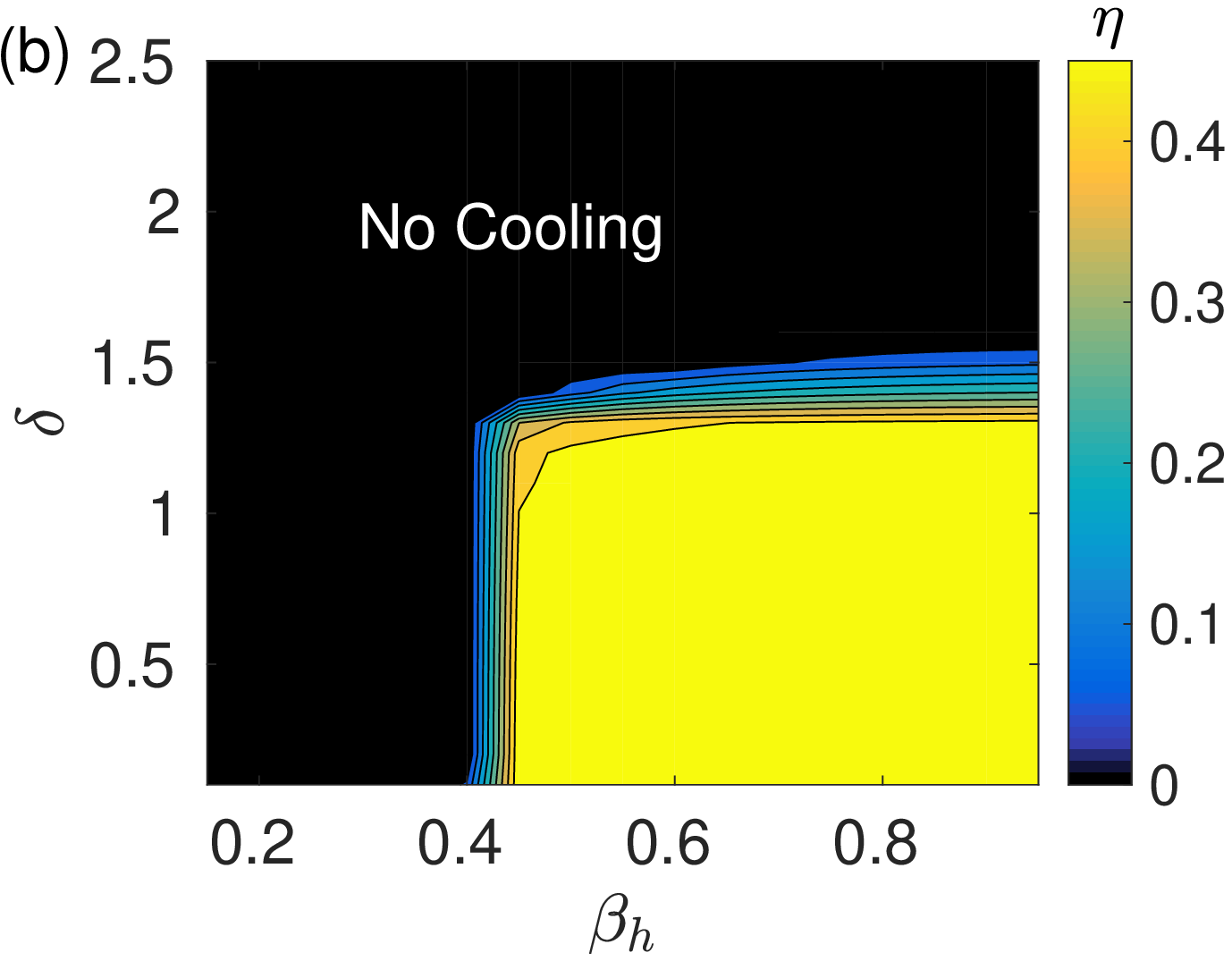}
\caption{ (a) Cooling window  and 
(b) corresponding cooling COP for the configuration described in Fig.~\ref{Figbox}.
}
\label{Figcool}
\end{figure}

In Fig.~\ref{Figcool}, we display the cooling current and the cooling COP for the model of Fig.~\ref{Figbox}.
The current  $J_c/C_3$ is dimensionless, with $C_3\propto \gamma^2$,  $\gamma$ as the system-bath interaction energy,
see text below Eq. (\ref{eq:Jc3}). To receive the current in physical units of e.g. J/sec, one should further
divide it by $\hbar$. 
We find that the system can act as a refrigerator within a certain domain:
According to the ideal case, Eq.~(\ref{eq:cool-op}), cooling takes place when
$\beta_h\geq \frac{\theta_c}{\theta_h} (\beta_c-\beta_w) + \beta_w$. In our parameters, this reduces to
$\beta_h \geq 0.9/3 + 0.1=0.4$. This estimate qualitatively agrees with simulations at small $\delta$.

The width parameter is important as well, and we find that refrigeration takes place
as long as $\delta\lesssim 1.6$. Note that when $\delta\geq1.8$,
 the functions $M_{c}(\omega)$ and $M_w(\omega)$ overlap leading to energy leakage directly from the work bath
to the cold environment.
Nevertheless, it is interesting to note that the hot and work reservoirs already
touch when $\delta \geq 2/3$, yet their overlap does not prohibit the cooling process.
Overall, temperature, $\theta_{\nu}$, and the width parameters $\delta_{\nu}$ intermingle
in the expression for cooling. As a result,
the cooling window depends on $\delta$ in a non-trivial manner.
As an additional comment, in Fig.~\ref{Figcool}, we consistently receive $\eta<\theta_c/\theta_w=0.5$.
However, when using a narrower window for the cold bath, e.g.,
$\delta_c\sim 0.1$, we find that in fact the cooling COP can {\it exceed} $1/2$,
yet  obviously it still lies below the Carnot bound.

\vspace{6mm}
\begin{figure}[htbp]
\hspace{4mm}
\includegraphics[width=12cm]{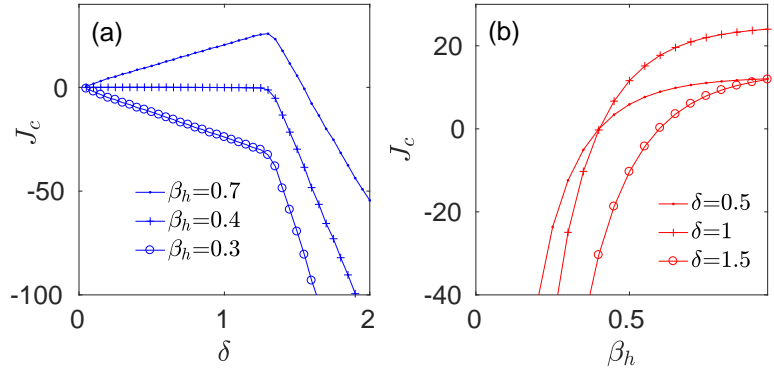}
\caption{Slices of cooling current from Fig.~\ref{Figcool}.
(a) $J_c$ as a function of $\delta$ for three  values of $\beta_h$.
(b) $J_c$ as a function of $\beta_h$ for three  values of $\delta$. 
Other parameters are the same as indicated in Fig.~\ref{Figbox}.
}
\label{Figcut}
\end{figure}
%


In Fig.~\ref{Figcut}, we display slices of the cooling current 
as a function of the width $\delta$ and temperature $\beta_h$, taken from the contour plot,  Fig.~\ref{Figcool}.
For small value of $\delta$ the cooling current grows linearly. However, at a certain point ($\delta=1.3$), 
when the cold and work reservoirs almost touch, the cooling current begins to drop with $\delta$, 
eventually missing the cooling operation altogether. 
The behavior of the cooling current with $\beta_h$ is monotonic.

\begin{figure}[h]
\hspace{4mm}
\includegraphics[width=13cm]{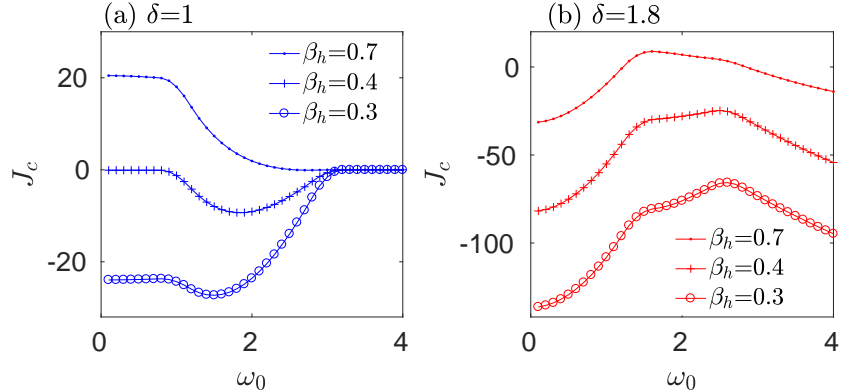}
\caption{Cooling current as a function of the qubit energy gap $\omega_0$ for
the parameters of Fig.~\ref{Figbox}-\ref{Figcool} for $\beta_h=0.3,0.4,0.7$, bottom to top,
and (a) $\delta=1$ , (b) $\delta=1.8$.
}
\label{Figw0}
\end{figure}

The role of a nonzero gap $\omega_0\neq 0$ on the operation of the qubit-QAR is presented in Fig.~\ref{Figw0},
using again parameters as in Fig.~\ref{Figbox}.
In panel (a), we set $\delta =1$. We find that within the cooling window, the performance of the QAR is intact for 
small $\omega_0$, but it deteriorates and eventually disappears once $\omega_0$ is comparable to differences of
the central frequencies, $\theta_{\nu}$.
Outside the cooling window, for large $\delta$, in panel (b) we reveal a non-trivial non-monotonic behavior of $J_c$ 
with $\omega_0$, with a large value of $\omega_0$ manifesting a cooling function that is missing in the degenerate case.
Overall, as long as $\omega_0\ll \delta$, it plays an insignificant role in the refrigeration behavior. 
Beyond that, it can introduce cooling, enhance or reduce performance, depending on the particular choice of parameters.

\section{Physical model}
\label{sec-physics}

Before concluding, we discuss physical models corresponding to the model Hamiltonian~(\ref{eq:Hs}) and the functions $M_{\nu}(\omega)$. 
We assume that the baths include collections of harmonic oscillators, 
$\hat H_{B,\nu}=\sum_{j}\omega_{j,\nu}\hat b_{j,\nu}^{\dagger}\hat b_{j,\nu}$, 
and that bath operators, which are coupled to the system, are of the form
\bea
\hat B_{\nu}=\sum_{j}\frac{\lambda_{j,\nu}}{\omega_{j,\nu}}\left(\hat b_{j,\nu}^{\dagger}+\hat b_{j,\nu}\right).
\eea
The two-time correlation functions are thus given by
\bea
M_{\nu}(t)=\langle \hat B_{j,\nu}(t)\hat B_{j,\nu}(0)\rangle= \sum_{j}\frac{\lambda_{j,\nu}^2}{\omega_{j,\nu}^2}
\left[ n_{\nu}(\omega_{j,\nu})e^{i\omega_{j,\nu}t}
+(n_{\nu}(\omega_{j,\nu})+1)e^{-i\omega_{j,\nu}t}
 \right],
\eea
with $n_{\nu}(\omega)=[e^{\beta_{\nu}\omega }-1]^{-1}$ as the Bose-Einstein distribution function. In the frequency domain,
\bea
M_{\nu}(\omega)= 2\pi\sum_{j}\frac{\lambda_{j,\nu}^2}{\omega_{j,\nu}^2}\left[ n_{\nu}(\omega_{j,\nu})\delta(\omega+\omega_{j,\nu})
+(n_{\nu}(\omega_{j,\nu})+1)\delta(\omega-\omega_{j,\nu})
 \right].
\eea
We now define the spectral density function, which is related to the density of states of the bath,
\bea
g_{\nu}(\omega)=2\pi \sum_{j}\frac{\lambda_{j,\nu}^2}{\omega_{j,\nu}^2}\delta(\omega-\omega_{j,\nu}),
\eea
and find that,
\bea
M_{\nu}(\omega)=
\left[n_{\nu}(\omega)+1\right]g_{\nu}(\omega) 
\label{eq:M}
\eea
with $g_{\nu}(\omega)$ 
analytically continued to the complete real axis as an odd function.
According to Eq.~(\ref{eq:M}), the function $M_{\nu}(\omega)$ is linearly related to the spectral density 
function $g_{\nu}(\omega)$ of the respective bath.
Thus, if we can engineer the density of states of the bath, or filter its frequencies, 
we can realize a structured model of the form~(\ref{eq:steps}).

Let us now also comment on the relation of the Hamiltonian~(\ref{eq:Hs}) 
to the nonequilibrium (multi-bath) spin-boson Hamiltonian. We begin with the additive model,
\bea
\hat H=\frac{\omega_0}{2}\hat \sigma_{z} + \frac{\Delta}{2} \hat \sigma_{x}
+ \sum_{\nu,j}\omega_{j,\nu}\hat b_{j,\nu}^{\dagger}\hat b_{j,\nu} 
+ \hat \sigma_z\sum_{\nu,j} \lambda_{j,\nu} \left(\hat b_{j,\nu}^{\dagger}+\hat b_{j,\nu}\right),
\label{eq:HSB}
\eea
and  transform it 
to the displaced bath-oscillators basis using the small polaron
transformation~\cite{Mahan}, $\hat H_p=\hat U^{\dagger}\hat H\hat U$,
$\hat U=e^{i\hat \sigma_z\hat \Omega/2}$,
\bea
\hat H_{p} = \frac{\omega_0}{2} \hat \sigma_z +
\frac{\Delta}{2} \left( \hat \sigma_+ e^{i\hat \Omega} + \hat \sigma_- e^{-i\hat \Omega} \right)
+\sum_{\nu,j}\omega_{j,\nu} \hat b_{j,\nu}^{\dagger}\hat b_{j,\nu}.
\label{eq:HSBs}
\eea
Here, $\hat \sigma_{\pm}=\frac{1}{2}(\hat \sigma_x\pm i \hat \sigma_y)$
are the auxiliary Pauli matrices, $\hat \Omega=\sum_{\nu}\hat \Omega_{\nu}$, 
$\hat \Omega_{\nu}=2i\sum_{j}\frac{\lambda_{j,\nu}}{\omega_{j}}(\hat b_{j,\nu}^{\dagger}-\hat b_{j,\nu})$.
It can be shown that under the so-called noninteracting blip approximation~\cite{Weiss,Legget,Dekker}, 
the population dynamics follows Eq.~(\ref{eq:popd})
with the time correlation function 
\bea
M_{\nu}(t)=\langle e^{i\Omega(t)}e^{-i\Omega(0)}\rangle= e^{-Q_{\nu}(t)}.
\label{eq:MNIBA}
\eea
The average is performed with respect to the initial, product thermal state of the baths.
The function $Q(t)=\sum_{\nu}{Q_{\nu}(t)}$ is complex,  with real
 and imaginary  components, $Q_{\nu}(t)=Q'_{\nu}(t)+iQ''_{\nu}(t)$,
\bea
Q''_{\nu}(t)& = & 2 \int_{0}^{\infty} \frac{g_{\nu}(\omega)}{\pi}\sin(\omega t)d\omega,
\nonumber\\
Q'_{\nu}(t)& = &2 \int_{0}^{\infty}\frac{g_{\nu}(\omega)}{\pi}[1-\cos(\omega t)] [1+2n_{\nu}(\omega)] d\omega.
\label{eq:Q}
\eea
The multi-bath spin-boson example clearly demonstrates that non-additivity of the interaction Hamiltonian 
embodies strong coupling effects.
Using Eq.~(\ref{eq:HSBs}) as a starting point, a second order time-evolution scheme with respect to $\Delta$ 
provides an equation of motion for the population dynamics in the form 
(\ref{eq:popd}), which is beyond second order in the system-bath interaction energy $\lambda$~\cite{segal-nicolin}. 
Nevertheless, to realize a QAR we need to design the function $M_{\nu}(t)$, which is
the central object in the expressions for the population and energy current.
For the spin-boson model, this function, 
[see Eqs.~(\ref{eq:MNIBA})-(\ref{eq:Q})], depends in a non-linear manner on the reservoirs' density of states, 
thus it is not immediately clear how to physically-rationally engineer it to receive the form~(\ref{eq:steps}).

\section{Summary and Outlook}

What is the smallest possible fridge? Using a quantum master equation in Lindblad form with additive dissipators,
it was argued in Ref.~\cite{PopescuPRL} that a three-level system (qutrit) is the smallest refrigerator, 
if each transition is thermalized independently.
Here, we demonstrate that in fact a qubit could serve as smallest refrigerator---if the three thermal 
reservoirs $c$, $h$ and $w$
couple in a non-additive manner to the qubit, to transfer energy in a cooperative manner.
Our formalism is consistent with classical thermodynamics, as it complies with the heat exchange fluctuation theorem.

In the ideal limit we receive refrigeration if
(i) the reservoirs are engineered, and are allowed to exchange energy within a highly restricted spectral window, 
(ii) a resonant condition $\theta_c+\theta_w=\theta_h$ is satisfied for the characteristic frequencies of the three reservoirs.
In this special situation,  we derived closed expressions for 
the cooling window, the efficiency of the refrigerator (characterized by its COP),
the maximal efficiency (bounded by the Carnot limit) and the maximum-power efficiency. We found that
a qubit-QAR with a non-additive dissipation form
can embody function that is prohibited under the weak system-bath coupling assumption.
We had further studied, analytically and numerically, the operation of the system beyond the ideal limit and found
that it can operate as a QAR for a fair range of parameters.

Throughout this study, we had assumed a resonance condition~(\ref{eq:resonant}) for the characteristic frequencies of
the three reservoirs, so as to optimize performance.
Nevertheless, it is imperative to realize that one could construct a qubit-QAR without structuring the work reservoir.
Going back to Eq.~(\ref{eq:JQARu}), we  evaluate the cooling current when the cold and hot FBCF take the form
of a Dirac delta function as in Eq.~(\ref{eq:Mdelta}), with $\theta_h>\theta_c>0$,
but the work reservoir's FBCF is simply a constant, 
$M_{w}(\omega)=\epsilon_w$ for $\omega\geq0$.
We then obtain a modified cooling window (compare to Eq.~(\ref{eq:etadelta})),
\bea
\theta_c\left(  e^{-\beta_c\theta_c} e^{-\beta_w(\theta_h-\theta_c)} - e^{-\beta_h\theta_h} \right) + 
\theta_c\left(  e^{-\beta_c\theta_c} e^{-\beta_h\theta_h} - e^{-\beta_w(\theta_c+\theta_h)}
 \right) \geq 0.
\eea
The new, last two terms correspond to the extraction of energy from the cold bath, assisted by the hot bath, to be dumped into
the work reservoir, and the reversed process.
We can reorganize the cooling condition as
\bea
e^{-\beta_c\theta_c} e^{-\beta_w(\theta_h-\theta_c)} - e^{-\beta_h\theta_h}(1+A)\geq 0,
\eea  
where $A\equiv e^{-\beta_w(\theta_c+\theta_h)}e^{\beta_h\theta_h} -e^{-\beta_c\theta_c}$.
After some manipulations, we arrive at
\bea
\frac{\theta_c}{\theta_h} \leq 
\frac{\beta_h-\beta_w}{\beta_c-\beta_w} - \frac{1}{\theta_h(\beta_c-\beta_w)}\ln(1+A).
\label{eq:cool-opN}
\eea
Since $A>0$, it is clear that the cooling window is reduced here relative to the case with a structured work 
bath,~(\ref{eq:cool-op}).
This new situation is significant: We do not place here stringent conditions on the work reservoir, which could in fact 
be featureless and still support cooling.  Furthermore, this scenario, where the work reservoir is allowed to provide or absorb 
both low and high frequencies, is obviously un-accounted for in the traditional
three-level weak-coupling setting where each bath excites a particular transition in a resonant manner (see Fig.~\ref{FigQAR})(a).
A strongly-coupled qubit-QAR thus offers new regimes
of operation, missing in multi-level, weakly coupled designs.

We examined an additive dissipation model and proved that it cannot support a QAR performance based on a qubit.
In contrast, a non-additive model, with the three reservoirs acting in a 
concerted manner, can achieve refrigeration. It is of an interest to examine an in-between model, which could be more
practical. For example, the cold and work reservoirs could couple strongly to build up a non-additive dissipator, 
but the hot bath would weakly-separately couple to the system, bringing in an additional, local dissipation term. 
This scenario could be treated using the reaction coordinate method~\cite{Gernot} or the polaron-transformed master
equation~\cite{Ren1,Ren2,Schallerpol1,Schallerpol2}, which can smoothly interpolate between additive and non-additive dissipation problems.

The description of quantum systems that are strongly coupled to
multiple thermal reservoirs poses a significant theoretical challenge. 
Markovian master equations of Lindblad form with additive (local) dissipators provide a consistent
thermodynamical description of observables~\cite{kos13}.
However, it is not yet established how to formulate
quantum thermodynamics beyond the weak coupling limit. 
Our work here exemplifies a strong coupling framework which is thermodynamically consistent.
It is of interest to compare our approach and predictions to other recent studies 
on strongly-coupled energy conversion devices~\cite{DavidS, Kosloff,Cao,Gernot,Jarzynski,esposito17,Nazir}, and further 
develop numerically exact techniques that could target such problems.
Understanding the role of both strong coupling and non-Markovianity of the reservoirs on the operation of 
driven and autonomous thermal machines remains a challenge for future work.

\begin{acknowledgments}
DS acknowledges support from an NSERC
Discovery Grant and the Canada Research Chair program.
The work  of AM was supported by the CQIQC summer fellowship at
the University of Toronto.
GS gratefully acknowledges financial support by the DFG (SCHA 1646/3-1, GRK 1558).
Discussions with Javier Cerrillo are gratefully acknowledged. 
\end{acknowledgments}

\renewcommand{\theequation}{A\arabic{equation}}
\setcounter{equation}{0}  

\section*{Appendix A: Cooling condition for a non-degenerate Qubit-QAR}

We investigate here the cooling performance of a qubit of a finite energy gap $\omega_0$.
For simplicity, we assume the following functions ($\omega>0$). 
The negative-frequency branch is decorated by detailed-balance (thermal) factors,
\bea
M_c(\omega)&=&\epsilon_c\delta(\omega-\theta_c),
\nonumber\\
M_h(\omega)&=&\epsilon_h \left[ \delta(\omega-\theta_h-\omega_0)+\delta(\omega-\theta_h+\omega_0)\right],
\nonumber\\
M_w(\omega)&=&\epsilon_w\delta(\omega-\theta_w).
\label{eq:Mapp}
\eea
Here, $\epsilon_{c,h,w}$ are dimensionless parameters. 
Since these parameters do not influence the cooling window and the cooling
efficiency, we ignore them below.
We maintain the resonant condition, 
$\theta_c+\theta_w=\theta_h$. We further assume that $\omega_0$ is much smaller than $\theta_{\nu},\nu=c,h,w$. 

One should note that the steady-state populations of the qubit are no-longer equal. We will begin by evaluating them from Eq.
(\ref{eq:popd}). The three-bath convoluted decay rate is
\bea
M(\omega_0)=\int_{-\infty} ^{\infty}d\omega_1M_c(\omega_1)\int_{-\infty} ^{\infty} d\omega_2  M_h(\omega_2)M_w(\omega_0-\omega_1-\omega_2).
\eea
We break the integral into four contributions. For $\omega_1\geq0$, $\omega_2\geq0$,
\bea
M(\omega_0)&=&\int_{0} ^{\infty}d\omega_1M_c(\omega_1)\int_{0} ^{\infty} d\omega_2  M_h(\omega_2)M_w(\omega_0-\omega_1-\omega_2) 
\nonumber\\
&=& 0,
\eea
since we assume that $\omega_0$ is sufficiently small, such that $\theta_c+\theta_h-2\omega_0> \theta_w$.
Similarly, when $\omega_1\leq0$, $\omega_2\leq0$,
\bea
M(\omega_0)&=&\int_{-\infty} ^{0}d\omega_1M_c(\omega_1)\int_{-\infty} ^{0} 
d\omega_2  M_h(\omega_2)M_w(\omega_0-\omega_1-\omega_2)
\nonumber\\
&=&0.
\eea
We receive a finite contribution when $\omega_1\geq0$  and $\omega_2\leq0$,
\bea
M(\omega_0)&=&\int_{0} ^{\infty}d\omega_1M_c(\omega_1)\int_{-\infty} ^{0} d\omega_2  M_h(\omega_2)M_w(\omega_0-\omega_1-\omega_2)
\nonumber\\
&=&\int_{0} ^{\infty}d\omega_1M_c(\omega_1)\int_{0} ^{\infty} d\omega_2  M_h(\omega_2)e^{-\beta_h \omega_2}M_w(\omega_0-\omega_1+\omega_2)
\nonumber\\
&=&
e^{-\beta_h(\theta_h-\omega_0)},
\eea
as well as for  $\omega_1\leq0$,  and $\omega_2>0$,
\bea
M(\omega_0)&=&\int_{-\infty} ^{0}d\omega_1M_c(\omega_1)\int_{0} ^{\infty} d\omega_2  M_h(\omega_2)M_w(\omega_0-\omega_1-\omega_2)\nonumber\\
&=&\int_{0} ^{\infty}d\omega_1M_c(\omega_1)e^{-\beta_c\omega_1}\int_{0} ^{\infty} d\omega_2  M_h(\omega_2)M_w(\omega_0+\omega_1-\omega_2)\nonumber\\
&=&
\int_{0} ^{\infty}d\omega_1M_c(\omega_1)e^{-\beta_c \omega_1}\int_{0} ^{\infty} d\omega_2  M_h(\omega_2)M_w(\omega_2-\omega_1-\omega_0)e^{-\beta_w(\omega_2-\omega_1-\omega_0)}\nonumber\\
&=&e^{-\beta_c\theta_c}e^{-\beta_w\theta_w}.
\eea
Therefore, we find that $M(\omega_0)=e^{-\beta_h(\theta_h-\omega_0)}+e^{-\beta_c\theta_c}e^{-\beta_w\theta_w}$.
%
%
Similarly,  $M(-\omega_0)=e^{-\beta_h(\theta_h+\omega_0)}+e^{-\beta_c\theta_c}e^{-\beta_w\theta_w}$.  
In steady state, the populations follow  $p_g=\frac{M(\omega_0)}{M(\omega_0)+M(-\omega_0)}$ 
and $p_e=\frac{M(-\omega_0)}{M(\omega_0)+M(-\omega_0)}$.
Our results reduce to $p_{g,e}=1/2$ when $\omega_0=0$.

We now turn to the expression for the cooling current, Eq.~(\ref{eq:Jc3}).
It is easy to calculate the integrals in this equation since there is only one  extra 
$\omega_1$ term, which is trivial to account for under
(\ref{eq:Mapp}). We evaluate the two integrals,
\bea
I_1&\equiv&-\int_{-\infty} ^{\infty} \int_{-\infty} ^{\infty}
d\omega_1 d\omega_2 \omega_1 M_c(\omega_1)M_h(\omega_2)M_w(\omega_0-\omega_1-\omega_2)
\nonumber\\
&=&\theta_c(e^{-\beta_c\theta_c}e^{-\beta_w\theta_w}-e^{-\beta_h\theta_h}e^{\beta_h\omega_0}) , 
\eea
and
\bea
I_2&\equiv&\int_{-\infty} ^{\infty} \int_{-\infty} ^{\infty}d\omega_1 d\omega_2 \omega_1 M_c(-\omega_1)M_h(-\omega_2) M_w(-\omega_0+\omega_1+\omega_2)
\nonumber\\
&=&\theta_c(e^{-\beta_c\theta_c}e^{-\beta_w\theta_w}-e^{-\beta_h\theta_h}e^{-\beta_h\omega_0}) .
\eea
In total, the cooling current is
\bea
J_c&=& -p_e\int_{-\infty} ^{\infty}\int_{-\infty} ^{\infty}
d\omega_1 d\omega_2 \omega_1 M_c(\omega_1)M_h(\omega_2)M_w(\omega_0-\omega_1-\omega_2)
\nonumber\\  
&+&p_g\int_{-\infty} ^{\infty}\int_{-\infty} ^{\infty}d\omega_1 d\omega_2 \omega_1 M_c(-\omega_1)M_h(-\omega_2) M_w(-\omega_0+\omega_1+\omega_2)
\nonumber\\
&=&\frac{\theta_c}{M(\omega_0)+M(-\omega_0)}
\left(e^{-\beta_c\theta_c}e^{-\beta_w\theta_w}+e^{-\beta_h\theta_h}e^{-\beta_h\omega_0}\right)
\left(e^{-\beta_c\theta_c}e^{-\beta_w\theta_w}-e^{-\beta_h\theta_h}e^{\beta_h\omega_0}\right)
\nonumber\\
&+&\frac{\theta_c}{M(\omega_0)+M(-\omega_0)}
\left(e^{-\beta_c\theta_c}e^{-\beta_w\theta_w}+e^{-\beta_h\theta_h}e^{\beta_h\omega_0}\right) 
\left(e^{-\beta_c\theta_c}e^{-\beta_w\theta_w}-e^{-\beta_h\theta_h}e^{-\beta_h\omega_0}\right )
\nonumber\\
&=&\frac{2\theta_c}{M(\omega_0)+M(-\omega_0)}\left(e^{-2\beta_c\theta_c}e^{-2\beta_w\theta_w}-e^{-2\beta_h\theta_h}\right). 
\eea
The cooling window,  $J_c\geq0$, precisely follows Eq.~(\ref{eq:etadelta})---obtained for the case $\omega_0=0$.

\renewcommand{\theequation}{B\arabic{equation}}
\setcounter{equation}{0}  

\section*{Appendix B: Un-attainability of cooling for continuous-Gaussian spectral functions.}

As we showed in the main text, functions $M_{\nu}(\omega)$  of restricted range  support the cooling function in a qubit-QAR when
the interaction of the qubit with the separate baths is made non-additive. 
We now examine an example with functions that are non-zero in the full range, without any filtering, and show that the system cannot act as a QAR.

We assume a Gaussian form for the FBCF, 
$M_{\nu}(\omega)=\sqrt{\frac{\pi}{E_r^{\nu}T_{\nu}}}\exp\left(-\frac{(\omega-E_r^{\nu})^2}{4E_r^{\nu}T_{\nu}}\right)$,
which describes bath-induced rates in the spin-boson model at high temperature. 
Specifically, performing a short time expansion of Eq.~(\ref{eq:Q}), we get 
$Q^{''}_{\nu}(t)=\frac{2}{\pi}t \int d\omega g_{\nu}(\omega)\omega$ and 
$Q^{'}_{\nu}(t)=  \frac{2T_{\nu}}{\pi}  t^2 \int d\omega  g_{\nu}(\omega)\omega$.
Defining $E_r^{\nu} \equiv  \frac{2}{\pi}\int d\omega g_{\nu}(\omega)\omega$, we arrive at
$Q^{''}_{\nu}(t)=E_{r}^{\nu} t$ and
$Q^{'}_{\nu}(t)=E_{r}^{\nu}T_{\nu} t^2$.
The energy scale $E_r^{\nu}$ is referred to as  the ``reorganization energy", and it represents
the strength of the system-bath interaction.  
The current from the cold bath is given by Eq.~(\ref{eq:JQARu}), which solves to
\bea
J_c&=& -\frac{1}{2}\int_{-\infty} ^{\infty} \int_{-\infty} ^{\infty}
d\omega_1 d\omega_2 \omega_1 M_c(\omega_1)M_h(\omega_2)M_w(-\omega_1-\omega_2)  
+\frac{1}{2} \int_{-\infty} ^{\infty}d\omega_1 d\omega_2 \omega_1 M_c(-\omega_1)M_h(-\omega_2) M_w(\omega_1+\omega_2)
\nonumber\\
&=&-\frac{1}{2}
\int_{-\infty} ^{\infty}\int_{-\infty} ^{\infty}d\omega_1 \omega_1 M_c(\omega_1) \int_{-\infty}^{\infty} 
d\omega_2 M_h(\omega_2) M_w(\omega_1+\omega_2)\left[ e^{-\beta_w(\omega_1+\omega_2)}- e^{-\beta_c\omega_1}e^{-\beta_h\omega_2} 
\right]\nonumber\\
&=&
4E_r^c{\pi}^{\frac{5}{2}}\frac{E_r^h(T_c-T_h)+E_r^w(T_c-T_w)}{(E_r^cT_c+E_r^hT_h+E_r^wT_w)^{\frac{3}{2}}}
\exp\left(-\frac{(E_r^c+E_r^h+E_r^w)^2}{4(E_r^cT_c+E_r^hT_h+E_r^wT_w)}\right).
\label{eq:gauss}
\eea
Since this expression is always negative, the system cannot act as a QAR.
We can rationalize this result as follows. Let us define an effective temperature
\bea
T^*\equiv \frac{E_r^cT_c+E_r^hT_h+E_r^wT_w}{E_r^c+E_r^h+E_r^w},
\eea
and the total interaction-reorganization energy, $E_r\equiv \sum_{\nu} E_r^{\nu}$.
We can then recast  $J_c$ of Eq. (\ref{eq:gauss}) as follows,
\bea
J_c\propto \frac{1}{\sqrt{T^{*}E_r }}\frac{\left(T_c-T^*\right)}{T^*} e^{-E_r/4T^*}.
\eea
Since $T^*>T_c$, cooling is prohibited. 
We now understand that, since  the reservoirs can absorb and emit all
frequency components, the system (qubit) can in fact be 
characterized by an effective interaction energy and an effective temperature,
the latter is always greater than $T_c$ due to the cumulative effect of the other baths. As a result,
it is impossible to extract energy from the cold bath. 
This argument suggests that finite-range hard-cutoff functions should be used to design a QAR, 
as we explain in the main body of the paper.


\end{document}